\documentclass[aps,prb,twocolumn,groupedaddress,showpacs,floatfix]{revtex4}
\usepackage{graphicx}
\usepackage{dcolumn}
\usepackage{bm}
\begin{document}


\title{Monte Carlo simulations of magnetic ordering in the fcc Kagom\'{e} lattice}


\author{V. Hemmati, M. L. Plumer, and J. P. Whitehead}
\affiliation{Department of Physics and Physical Oceanography,
Memorial University, St. John's, NL, Canada, A1B 3X7}

\author{B. W. Southern}
\affiliation{Department of Physics and Astronomy, University of Manitoba, Winnipeg, MB, R3T 2N2, Canada}

\date{\today}

\begin{abstract}
Monte Carlo simulation results are reported on magnetic ordering in ABC stacked Kagom\'{e} layers with fcc symmetry for both XY and Heisenberg models which include exchange interactions with the eight near-neighbors.  Well known degeneracies of the 2D system persist in the 3D case and analysis of the numerical data provides strong evidence for a fluctuation-driven first-order transition to well-defined long-range order characterized as the layered $q=0$ (120-degree) spin structure.  Effects of varying the inter-layer coupling are also examined. The results are relevant to understanding the role of frustration in IrMn$_3$ alloys widely used by the magnetic storage industry as thin-films in the antiferromagnetic pinning layer in GMR and TMR spin valves.   Despite the technological importance of this structure, it has not previously been noted that the magnetic Mn-ions of fcc IrMn$_3$ form Kagom\'{e} layers.
\end{abstract}


\maketitle
\section{INTRODUCTION}
The phenomenon of pinning the magnetization direction of a thin-film ferromagnet due to coupling with an adjacent thin-film antiferromagent(AF) is key to spin-valve based magnetic transducer technology.\cite{exbias,ogrady10} Understanding exchange bias at the microscopic level has progressed substantially in recent years and suggests that a mechanism to stabilize domains in the surface layers of the AF is essential.\cite{exbias2} Nearly degenerate energy states of the surface spin structure, such as found in geometrically frustrated AFs,  can facilitate such domain formation.\cite{lhotel11}  Although there are many features of the Ir-Mn compounds, such as high N\'{e}el temperatures ($T_N \sim 900 ~K$) which have contributed to their being  one of the most commonly used materials for the pinning layer in mass produced spin valves, it has been argued that the spin frustration realized in the parent fcc compound IrMn$_3$ is relevant.\cite{tsunoda10} IrMn$_3$ is in a class of magnetic compounds having the CuAu$_3$ crystal structure which can be described as 2D triangular planes ABC stacked along $\langle 111 \rangle$ axes.\cite{kren66}  However, it has not been previously noted that the magnetic Mn ions reside on sites in these planes that form the Kagom\'{e} structure.  The present work uses Monte Carlo (MC) simulations to explore the spin ordering and phase transitions with near-neighbor (NN) exchange interactions of such a 3D system, which we call the fcc Kagom\'{e} lattice.

Heisenberg or XY spins on the 2D lattice formed from corner sharing triangles (see Fig. 1) exhibit a high degree of degeneracy in the case of near-neighbor exchange interactions where the only requirement is that the sum of spins on each triangle be zero (thus forming the 120$^0$-spin structure).\cite{lacroix11,chalker92}  The classical Heisenberg system shows entropy-driven planar spin nematic ``order from disorder'' at $T=0$ and has extensive entropy. The expected zero temperature limit of the specific heat, $C_V = \frac{11}{12} ~ k_B$ per spin, has been verified by Monte Carlo simulations.\cite{chalker92,zhitomirsky08} Long range ground state N\'{e}el order has been demonstrated by adding further-neighbor exchange interactions in the Heisenberg case giving rise to so-called $q=0$ (see Fig. 1) or $\sqrt{3} \times \sqrt{3}$ spin structures.\cite{harris92} The XY spin model on the 2D Kagom\'{e} has received much less attention, likely due to the lack of relevant experimental systems.   However, it has been argued that at $T=0$ the spin order should be that of the three-state Potts model,\cite{huse92} which is known to exhibit a weak first order transition at a non-zero temperature in 3D.\cite{plumer93} 

\begin{figure}[ht]
\includegraphics[width=0.3\textwidth, angle=90]{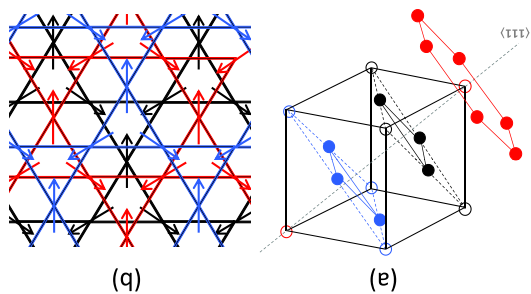}

\caption{(color online). (a) The fcc Kagom\'{e} lattice (solid circles) formed by ABC stacking of triangular layers (solid and open cirlces). In the case of IrMn$_3$, solid circles represent magnetic Mn-ions and open circles denote non-magnetic Ir-ions. (b) Projection of the $q=0$, 120$^0$ spin structure with ABC stacking along the $\langle 111 \rangle$ direction (out of the page) where colors distinguish the layers.}
\end{figure}

Three-dimensional structures are formed from weakly coupled ABC stacked Kagom\'{e} layers of magnetic ions in the family of compounds with rhombohedral symmetry known as the jarosites, AB$_3$(SO$_4$)(OH)$_6$, where a variety of experimental results suggest long range spin order of the $q=0$, 120$^0$-type, below temperatures in the range of 1 - 60 K.\cite{wills01,mendels11}  Stacked Kagom\'{e} layers have recently been investigated in an Fe-based metallo-organic compound which exhibits spin dynamics driven by frustration-induced domain walls.\cite{lhotel11} Numerical simulations of the magnetic phase transitions in these systems have not been reported.  Another 3D example is the hyperkagom\'{e} spin lattice, derived from the pyrochlore structure by removing 1/4 of the magnetic sites, which has been shown to exhibit the degeneracy of the 2D Kagom\'{e} lattice and MC simulations indicate a first order transition to octupolar ordering at very low temperature.\cite{zhitomirsky08} 

In addition to the very significant effort to understand the exchange bias mechanisms relevant to Ir-Mn thin films at the macroscopic level (see, e.g., Ref. [\onlinecite{ogrady10}] for a review), there have also been a number of works devoted to investigations of the antiferromagnetic spin order in bulk Ir-Mn (and sister) alloys at the microscopic level.  Early neutron diffraction results on samples with various relative concentrations of Ir in the disordered form of I$_r$Mn$_{1-x}$ alloys exhibit a simple two-sublattice magnetic order which suggest a peak in $T_N$ for values of $x$ close to 0.25.\cite{yamaoka74} The magnetic structure of IrMn$_3$ was found to possess the long-range 120$^0$-type spin order below $T_N  \simeq ~ 960 K$, referred to as the T1 state, which is equivalent to an ABC stacking of $q=0$ ordered Kagom\'{e} planes.\cite{tomeno99}  This magnetic structure had previously been proposed for the ordered phases of RhMn$_3$ and MnPt$_3$ alloys.\cite{kren66} This conclusion is also consistent with  first principles electronic structure calculations which, in addition, suggests that multiple-q spin stuctures are stabilized in the disordered alloys.\cite{sakuma03}  Most relevant to the present work involves a simulation of the spin structure of ordered IrMn$_3$ using the stochastic Landau-Lifshitz-Gilbert equations with a Heisenberg spin hamiltonian which includes a local $\langle 111 \rangle$ anisotropy term.\cite{szunyogh09}  Such simulations are essentially equivalent to the MC approach. Model parameters, which included a large number of exchange interactions up to 10 $\AA$ between neighbors, were determined from first-principles electronic structure calculations. The $q=0$ (T1) spin structure was verified and simulations of the order parameter vs temperature yielded the estimate $T_N \simeq  1145 ~  K$. 
\begin{figure}[ht]
\includegraphics[width=0.25\textwidth]{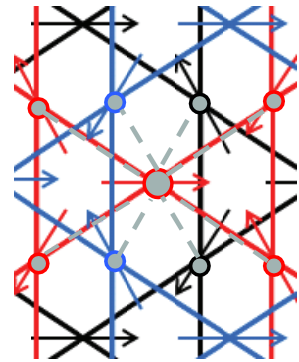}
\caption{ (color online). Illustration of the eight near-neighbors of the fcc Kagom\'{e} lattice projected onto the (111) plane (see Fig. 1).  Four neighbors are in the (red) plane, and two each are in adjacent planes (black and blue). }
\end{figure}
In the present work we examine the nature of the magnetic order, ground-state degeneracies and phase transitions in the fcc Kagom\'{e} Heisenberg and XY models with exchange interactions between the 8 NN shown in Fig. 2.  This is achieved through extensive ground-state as well as Metropolis MC simulations.  Results are presented for the temperature dependence of the internal energy, specific heat, order parameter and susceptibility all of which show clear indications of a phase transition to $q=0$ spin order for both models.  Analysis of the energy histograms as well as finite-size scaling of the specific heat and Binder cumulant indicate a strong first-order transition in the XY case but only weakly so for the Heisenberg model.  We propose a model of the ground state degeneracy involving planes of defects which appears to explain the multiple values of the sub-lattice magnetization found from the simulations at low T. The model also explains the differences between MC simulation results performed as heating, cooling or independent temperature runs. Additional MC simulation results are discussed for cases with weaker inter-layer exchange coupling which show the expected decrease in T$_N$.   The remainder of this paper is organized as follows.  In Sec. II, the model hamiltonian is presented as well as details on how the simulations were performed.  In Sec. III, the main results are shown for the temperature dependence of the various thermodynamic quantities.  Finite-size scaling analysis results are also discussed. This is followed by a description of our model of degeneracies.  Simulation results with weaker inter-layer coupling are presented in Sec. IV, followed by Sec. V where we discuss our results and directions for future simulations.   
\section{MODEL AND SIMULATIONS}

Monte Carlo simulations were performed using the standard Metropolis algorithm on $L\times L$ Kagom\'{e} planes, ABC stacked with $L$ layers. Periodic boundary conditions were used on lattices with L=12, 18, 24, 30, 36, and 60.  Between 10$^5$ and 10$^7$ MC steps (MCS) were used, with the initial ten percent discarded when calculating thermodynamic averages.  Only NN exchange interactions were included in the simulations, as defined by the following Hamiltonian   

\begin{eqnarray}
\mathcal{H} =   J  \sum_{i<j}^{\rm{intra-plane}} {\bf S}_i \cdot {\bf S}_j +  J'  \sum_{i<j}^{\rm{inter-plane}} {\bf S}_i \cdot {\bf S}_j. 
\end{eqnarray}
where the sum is over NN lattice sites, $J \equiv 1$ represents AF coupling to the four in-plane NN's and $J' > 0$ couples to the four NN sites in adjacent planes, as in Fig. 2. Most of the results below focus on the case of the fcc lattice stucture where $J'=J$, giving the full eight NN interactions. Results are also presented for systems with weaker inter-plane interactions, $J' < J$.  Simulations were performed assuming both XY and Heisenberg spin degrees of freedom.  In the XY case, spins were assumed to lie in plane.  

Preliminary to performing simulations on the stacked Kagom\'{e} lattice, extensive checks of the computer code were performed against published results for related systems.  The ground state spin configurations and energies for a 2D Kagom\'{e} model which also included second and third neighbor exchange interactions reproduced results in Ref. \onlinecite{harris92} for the boundaries between the $q=0$ and $\sqrt{3} \times \sqrt{3}$ phases.  MC simulations on the 2D system for the specific heat at low temperature gave results consistent with those in Ref. \onlinecite{zhitomirsky08} with NN exchange interations only.  Additional simulations were made on the ordinary Heisenberg fcc AF (ABC stacked triangular layers) where the temperature and first-order nature of the transition, based on the specific heat, were found to agree with previous results.\cite{diep89} 

\section{FCC KAGOM\'{E} LATTICE }

\subsection{Order of the transition}

A focus of these simulations was to establish the nature of the transitions to long range magnetic order in the XY and Hesienberg models on the fcc Kagom\'{e} lattice through evaluations of the energy and specific heat.  Fig. 3 shows the energy vs temperature of these two models for the case of $L=24$.  Three types of simulation results are shown.  In the case labelled cooling, the first simulation was run starting at a high temperature with a   random initial spin configuration.  Subsequent runs were made at decreasing temperatures where the initial spin configuration at each temperature was taken to be the final spin state of the previous run.  A corresponding technique was used for heating runs, starting from the $q=0$ ground state.  In these cases, thermal averaging was performed using 10$^6$ MCS on a single CPU. We also performed independent temperature simulations where a new random initial state was assumed at each value of $T$.  For this purpose, simulations at each temperature could each be performed on multiple, separate CPU and 10$^7$ MCS were used for averaging in these cases.  The results show a clear indication of phase transition for both models and the discontinuity close to $T_N=0.760$ for the XY model indicates it is first order for this two-component spin system.   In the Heisenberg case, the order of the transition close to $T_N=0.476$ is less clear and is further investigated below. These results can be compared with the first order transition estimated to be at $T_N=0.446$ in the ordinary fcc Heisenberg AF which involves collinear long range magnetic order.\cite{diep89}    The fact that the three types of simulations result in identical energy curves will be relevant to the discussion below of degenerate spin states. 

\begin{figure}[ht]
\includegraphics[width=0.4\textwidth]{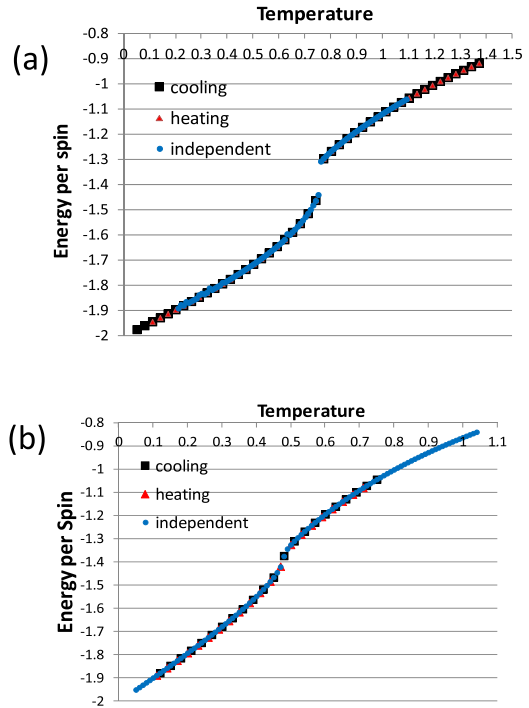}
\caption{ (color online). Energy of the fcc Kagom\'{e} (a) XY  and (b) Heisenberg models for $L=24$ with cooling, heating and independent temperature simulations. }
\end{figure}

Energy correlations were used to calculate the specific heat for the two models, shown in Fig. 4, using the same simulation runs as for the energy.  Well defined peaks occur at temperatures corresponding to the features in the energy plots of Fig. 2 and again there is no difference in results obtained from the three types of simulations. The curves also appear to suggest a zero temperature limit for the specific heat of close to 1/2 k$_B$ per spin in the XY case and close to 1 for the Heisenberg model.  The latter result is consistent with the value $\frac{11}{12} ~ k_B$ per spin known to occur in the 2D Heisenberg system.\cite{chalker92,zhitomirsky08} Such analysis would also suggest that the corresponding value for the XY model should be $\frac{5}{12} ~ k_B$ per spin, again consistent with our results.  Further indication that the degeneracies present in two dimensions persist for the 3D fcc Kagom\'{e} lattice are explored below.

\begin{figure}[ht]
\includegraphics[width=0.4\textwidth]{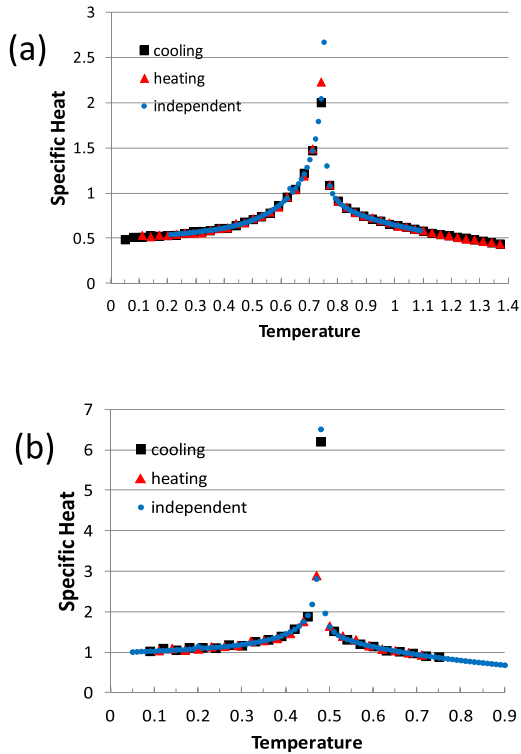}
\caption{ (color online). Specific heat of the fcc Kagom\'{e} (a) XY  and (b) Heisenberg models for $L=24$ with cooling, heating and independent temperature simulations. }
\end{figure}

Additional simulations were performed to explore finite size effects on the order of the phase transition for the Heisenberg case.  An indication that this transition is weakly first order in the thermodynamic limit is suggested by the clearer discontinuity in the energy shown in Fig. 5 for the larger lattice $L=36$ using 10$^7$ MCS for averaging and independent temperature runs.  Further evidence of the first order nature is also provided by the energy histograms shown in Fig. 6 which illustrate a discontinuous jump in the energy minimum as the temperature varies only slightly around $T_N$ for both the XY and Heisenberg models.\cite{challa86}  These histogram data also provide an accurate estimate of the respective transition temperatures corresponding to the value of $T$ which exhibits a double peak structure. Finally, the Binder energy cumulant\cite{challa86} was also calculated for the Heisenberg model using the larger lattice sizes with 10$^7$ MCS for averaging, as shown in Fig. 7.  The minimum exhibits clear finite-size volume scaling; however, its extrapolated value for $L \rightarrow \infty$ appears close to 2/3 expected of a continuous transition but our data are not accurate enough to make a conclusion on this point.  It is noteworthy that for the frustrated stacked triangular AF, the extremely weak nature of its first order transition was confirmed by numerical simulations only after some twenty years of study by a large number of groups.\cite{plumer94,ngo08}
\begin{figure}[ht]
\includegraphics[width=0.4\textwidth]{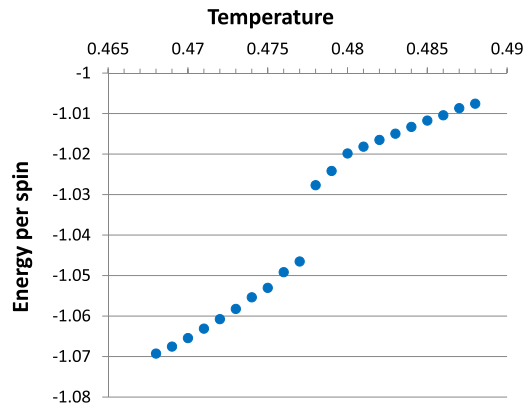}
\caption{ (color online). Energy of the fcc Kagom\'{e} Heisenberg model for $L=36$ with independent temperature simulations. }
\end{figure}
\begin{figure}[ht]
\includegraphics[width=0.4\textwidth]{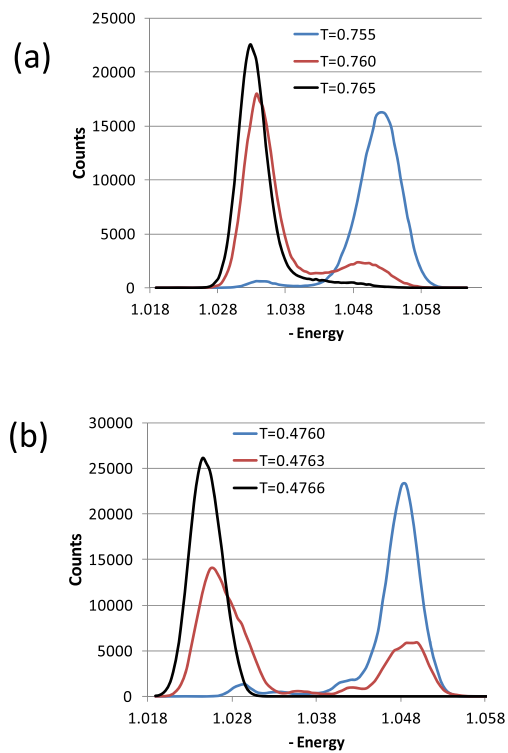}
\caption{ (color online). Energy histograms for the fcc Kagom\'{e} (a) XY  with $L=24$ and (b) Heisenberg models with L=60.   }
\end{figure}
\begin{figure}[ht]
\includegraphics[width=0.4\textwidth]{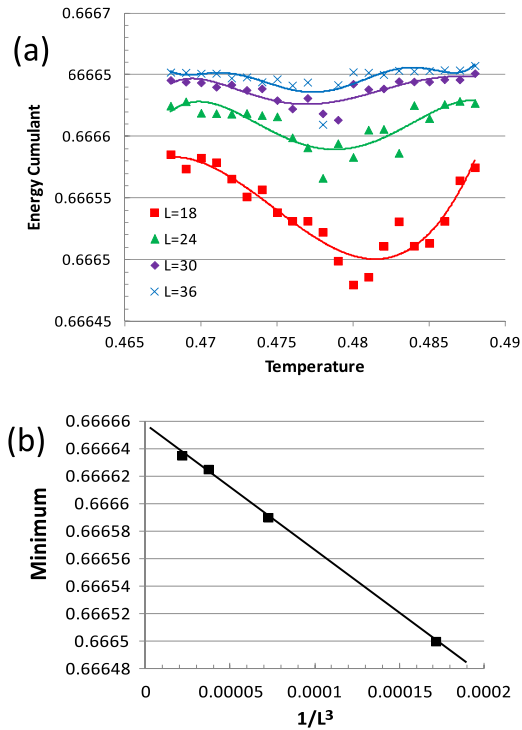}. 
\caption{ (color online).(a) Binder energy cumulant of  the fcc Kagom\'{e} Heisenberg model using independent temperature simulations. (b) Scaling of the minimum value vs inverse volume.}
\end{figure}

\subsection{Order parameter and susceptibility}

The $q=0$ ground state magnetic structure of the 2D Kagom\'{e} lattice is defined by the spins on each triangle being 120$^0$ apart.  This rule can also be satisfied in the ABC stacked fcc structure with eight NN spins, as shown in Fig. 2.  We have verified this to be the ground state for the 3D system and it is assumed here that this is the long range ordered state that occurs below the transition temperatures in the XY and Heisenberg cases considered in the previous section. The order parameter (OP) is defined through the three interpenetrating ferromagnetically aligned sublattices, ${\bf M}_\eta$ ($\eta = 1,2,3$)  associated with the 120$^0$ spin structure  
\begin{eqnarray}
M_t = (3/N) \{\langle M^2_1 + M^2_2 + M^2_3 \rangle/3\}^\frac{1}{2}
\end{eqnarray}
where $N$ is the number of sites and 
\begin{eqnarray}
M^2_\eta = (M^x_\eta)^2 + (M^y_\eta)^2 + (M^z_\eta)^2
\end{eqnarray}
with
\begin{eqnarray}
M^\sigma_\eta = \sum_i S^\sigma_{i,\eta},
\end{eqnarray}
where $\sigma = x,y,z$.  In the XY model, only $x$ and $y$ spin components are considered.

Fig. 8 shows the temperature dependences of the calculated OPs for the XY and Heisenberg models with $L=24$ for cooling, heating and independent temperatures simulations using 10$^6$ MCS for averaging. The curves obtained by heating the system from its 3D $q=0$ ordered state show full saturation of the OPs for $T \rightarrow$ 0 and smooth, monotonically decreasing functions as $T$ increases.  The curves for the cooling process are also smooth but do not saturate at low $T$.  The independent temperature runs are not monotonic and the OP value jumps between a number of values that mostly lie between the heating and cooling curves.   These results, together with the fact there is no difference in energy between these three types of simulations shown in Fig.3, indicate there are multiple degeneracies associated with the $q=0$ magnetic structure of the fcc Kagom\'{e} lattice.  These points are explored further in the next section.
\begin{figure}[ht]
\includegraphics[width=0.40\textwidth]{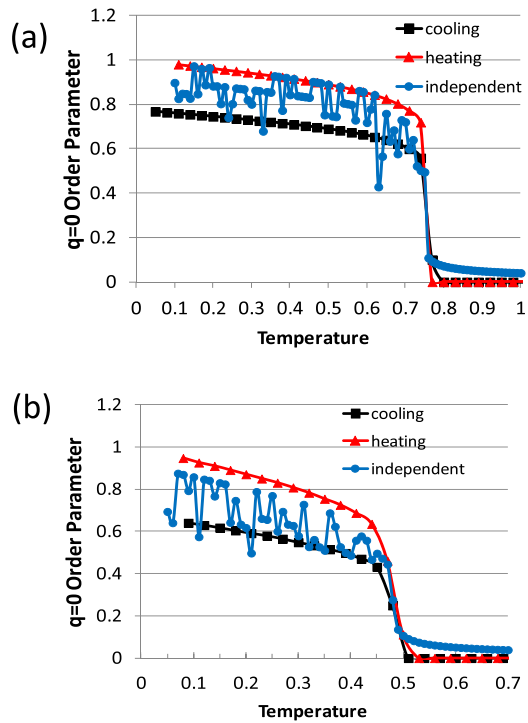}. 
\caption{(color online). $q=0$ order parameters for the (a) XY and (b) Heisenberg models with $L=24$ from cooling, heating and independent temperature simulations.}
\end{figure}

Spin correlations were also studied through the OP susceptibility response function defined by
\begin{eqnarray}
\chi = (\langle M_t^2\rangle - \langle |M_t| \rangle ^2)/(k_BT).
\end{eqnarray}
Fig. 9 shows cooling and heating runs only which are not identical due to the degeneracies illustrated in Fig. 8.  Independent temperature runs were also performed (not shown) and show large fluctuations due to the jumping between degenerate spin states (Fig. 8). The peaks in these curves occur near 0.76 and 0.47 for the XY and Heisenberg cases, respectively, are consistent with the transition temperature estimated based on the energy and specific heat anamolies, providing further evidence that the fcc Kagom\'{e} lattice indeed exhibits long range $q=0$ type spin order.  

\begin{figure}[ht]
\includegraphics[width=0.40\textwidth]{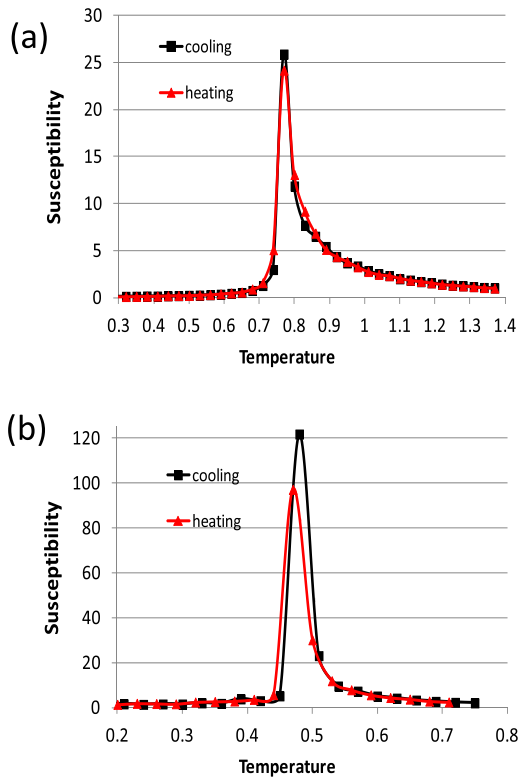}. 
\caption{ (color online). Susceptibilities corresponding to the $q=0$ order parameter for the (a) XY and (b) Heisenberg models with $L=24$ from cooling and heating simulations.}
\end{figure}

\subsection{Spin degeneracies}

In order to further investigate the impact of the $q=0$ spin configuration degeneracies associated with NN exchange interactions on the stacked Kagom\'{e} lattice, the three individual sub-lattice ferromagnetic OPs (M$_1$, M$_2$, and M$_3$) were calculated . Fig. 10 illustrates three example results of cooling runs performed on the XY model using three different random initial spin configurations.  In each case, one of the OPs saturates at $T=0$ to the expected value of 1/3 and the other two OPs both tend toward the same smaller values, approximately given by 0.22, 0.19, and 0.25.  Additional simulations starting with different random initial spin configurations exhibit other saturation values of the OPs although none are less than about 0.16.  Similar results were found for the Heisenberg model.

\begin{figure}[ht]
\includegraphics[width=0.4\textwidth]{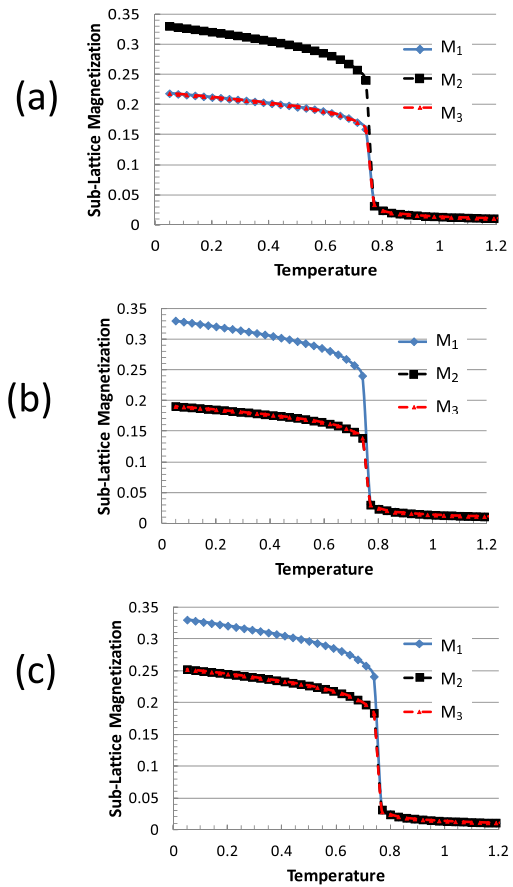}. 
\caption{(color online). The three ferromagnetic sub-lattice order paramaters associated with the $q=0$ spin state corresponding to three different cooling simulations, (a), (b), and (c), of the XY model with $L=24$.}
\end{figure}

\begin{figure}[ht]
\includegraphics[width=0.3\textwidth]{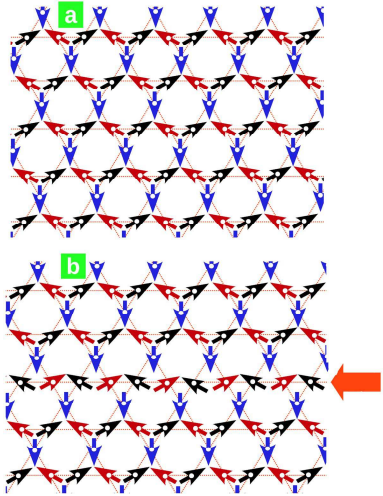}. 
\caption{ (color online). The three ferromagnetic sub-lattice spins shown as red, blue and black. (a) Perfect $q=0$ spin state. (b) Illustration of defects along the indicated horizontal line where two of the three ferromagnetic sub-lattice spins (black and red) are rotated by 120$^0$.}
\end{figure}

These results can be understood by the observation that there is no change in energy with an interchange of the direction of two sub-lattice spins along a line in 2D or in a plane in 3D. This is equivalent to rotating the spins along a line (or plane) by 120$^0$.\cite{chalker92,mendels11} The 2D case is illustrated in Fig. 11.  The possible values of the saturated sub-lattice magnetization due to this effect can be enumerated by considering all possible defect lines/planes of sub-lattice switching, and is given by  

\begin{eqnarray}
M_{\eta}=\frac{\sqrt{\left ( \frac{1}{4}~L^{3}-\frac{3}{2}~n \right )^{2}+\left (\frac{\sqrt{3}}{2}~n   \right )^{2}}}{\frac{3}{4}~L^{3}} 
\end{eqnarray}
where $n$ is the number of spins making the switch (e.g., between sub-lattices 1 and 2).  In 2D the smallest deviation from full saturation of a sub-lattice occurs if there is only one row of switched spins so that $n=L/2$.  In 3D, $k$ planes of switching involves $k(L/2)(L/2)$ spins up to a maximum where half of the population switches, $n=\frac{1}{8}L^3$ (where the number of spins on each sub-lattice is $\frac{1}{4}L^3$).  This means that the for the fcc Kagom\'{e} lattice the OP (at $T=0$) of each sub-lattice will lie in the range [$\frac{1}{6},\frac{1}{3}$] in the case of NN exchange interactions. The possible saturation values for the two switched sub-lattice OPs in the case $L=24$ are shown in Fig. 12 and are consistent with the values found in the simulation results of Figs. 8 and 10. 

\begin{figure}[ht]
\includegraphics[width=0.4\textwidth]{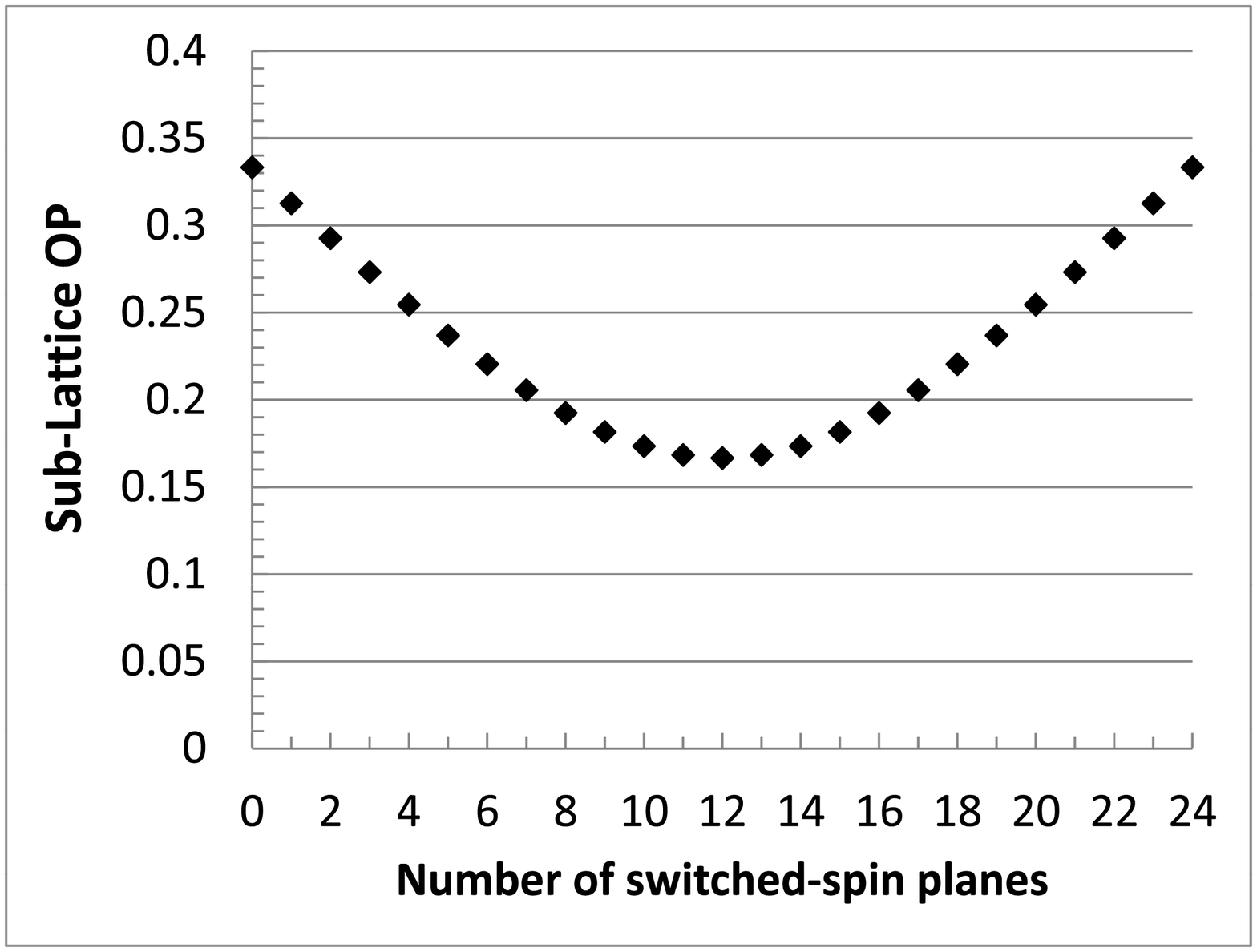}. 
\caption{ (color online). Possible values of the sub-lattice order parameter due to switched spins in the case of $L=24$. }
\end{figure}

Note that in each run, one of the sub-lattices undergoes no switching.  It is possible that in different parts of the lattice different sub-lattices will take the role of the one that does not switch.  Planes of switched spins formed in the cooling and heating processes usually are parallel; however, in some cases the planes may intersect each other. The spin configurations at these intersections impose an extra energy on the system. In a slow cooling process, these costly spin-configurations at the intersections of the planes are eliminated by thermal relaxation, and finally all the switched planes are aligned parallel. It may be possible to trap these costly spin-configurations in a fast cooling process and values of sublattice magnetization not predicted by Eq. (6) would occur.

\section{INTER-LAYER COUPLING}

In addition to bulk materials which exhibit fcc symmetry, magnetic compounds composed of ABC stacked Kagom\'{e} planes which are weakly exchange coupled are of interest (as noted in the Introduction).   We examine here the dependence of the transition temperature of both XY and Heisenberg models on the inter-layer exchange coupling $J'$ through simulations of the specific heat on $L=24$ systems with periodic boundary conditions and using 10$^7$ MCS for averaging.  Values $0.05 \leq J' \leq 0.95$ in steps of 0.05, with intra-plane exchange $J=1$,  were considered in independent temperature simulations distributed over 400 CPUs.

Example results for a few values of $J'$ are shown in Fig. 13 and a summary of the dependence of $T_N$ on $J'$ is given in Fig. 14.  The absence of a detectable peak in the specific heat at the smallest $J'$ over this range of $T \gtrsim 0.05$ is consistent with the predicted lack of long range spin order in 2D.  We find that the specific heat tends to the $T \rightarrow 0$ values observed for the fcc case (Fig. 4), independent of $J'$.  This result again suggests that NN inter-layer coupling does not alter the nature of the spin degeneracies of the 2D system.   The general shapes of these curves are consistent with previous investigations of quasi-2D systems.\cite{liu89,yasuda05}    

\begin{figure}[ht]
\includegraphics[width=0.40\textwidth]{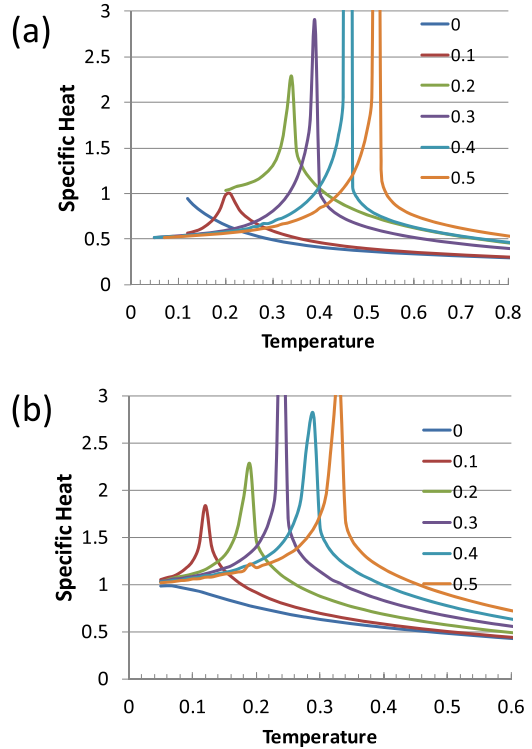}. 
\caption{ (color online). Example specific heat simulation results for interlayer coupling J$'<$1 as indicated for the (a) XY and (b) Heisenberg models from independent temperature simulations with $L=24$.}
\end{figure}

\begin{figure}[ht]
\includegraphics[width=0.40\textwidth]{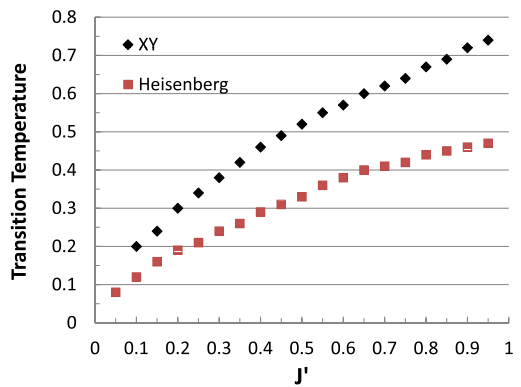}. 
\caption{ (color online). Transition temepratures for the XY and Heisenberg models with interlayer coupling J$'>$1.}
\end{figure}

\section{SUMMARY AND CONCLUSIONS}

A signifcant conclusion from the results presented in this work is that the fcc Kagom\'{e} spin lattice with NN exchange interactions ($J$) exhibits long range magnetic order of the $q=0$ (120$^0$) type through first order phase transitions at temperatures 0.760 $J$ and 0.476 $J$ for XY and Heisenberg models, respectively. In the Heisenberg case, finite-size scaling of the energy cumulant, along with other results, suggest that the first-order nature of the transition is very weak.  For both of these models, mean field theory would predict a continuous phase transition.  Spin configuration degeneracies well known in the 2D system as lines of defects appear to persist in the 3D case as planes of defects and lead to multiple (but well defined at $T=0$) possible values of the sub-lattice magnetization order parameters. These observations suggest that the transitions are driven by the order-by-disorder phenomenon. Additional simulations of ABC stacked Kagom\'{e} planes with weaker inter-layer coupling, $J'$,  exhibit similar types of transitions and degeneracies, with $T_N$ decreasing monotomically to zero as $J' \rightarrow 0$.

These results are relevant to a number of experimental systems.  Fcc IrMn$_3$ is one of the Ir-Mn compounds well known in the magnetic recording industry as useful for the pinning AF layer in spin valve devices.  Previous neutron scattering experiments have confirmed that this and sister compounds show long-range $q=0$ spin order (known in the industry as T1 magnetic order).  The pinning of the magnetization in a ferromagnetic layer through exchange coupling with an adjacent AF is believed to be due to the presence of defects and domains.  The results of the present work  emphasize that geometrical frustration is useful in generating these spin defects in compounds with the fcc Kagom\'{e} spin structure, like IrMn$_3$.  

This work represents a prelimnary investigation of a number of problems related to both fundamental aspects of geometrical frustation as well as exchange pinning.  We have already performed initial simulations on the behaviour of the fcc Kagom\'{e} system in an applied magnetic field.  Such studies previously revealed a tricritcal point in the case of the stacked triangular lattice and further illuminated the nature of the first order transition for that lattice structure.\cite{plumer93} Another extension of the present work is to examine the impact of the single-ion anisotropy noted in Ref. [\onlinecite{szunyogh09}], as well as Dzyaloshinskii-Moriya interactions,\cite{szunyogh11} on the nature of the phase transitions in 3D.  MC simulations of thin film ABC stacked Kagom\'{e} systems which include anisotropy as well as surface effects (on both exchange and anisotropy) would be useful to study surface spin-states and their possible impact on defects and domain formation.  Extensions to include also a ferromagnetic layer with dipole interactions would reveal more about the microscopic mechanisms important for exchange pinning for IrMn$_3$.\cite{ali03} Investigations of  spin waves in stacked Kagom\'{e} systems\cite{harris92} would compliment previous studies of the layered triangular AF\cite{meloche06}and would provide insight into the mechanism relevant to high frequency spin-valve sensor response.

\section{Acknowledgments}
This work was supported by the Natural Science and Engineering Research Council of Canada (NSERC) and the Atlantic Computational Excellence Network (ACEnet).  BWS thanks Memorial University and ACEnet for their hospitality.

\end{document}